\documentclass[11pt]{amsart}

\usepackage[ibidtracker=false,uniquename=false,giveninits=true,terseinits=true,backend=biber, sortcites=true]{biblatex}
\usepackage{float}
\usepackage{graphicx}
\usepackage{todonotes}
\usepackage{subcaption}
\usepackage{amsmath}
\usepackage{amsthm}
\usepackage{amssymb}
\usepackage{algorithm}
\usepackage[noend]{algorithmic}
\usepackage[foot]{amsaddr}
\usepackage[misc]{ifsym}
\usepackage{enumitem}
\usepackage{geometry}
\usepackage[hidelinks]{hyperref}

\renewbibmacro{in:}{}
\AtEveryBibitem{
	\clearlist{language}
}

\setlist{leftmargin = 0pt}
\newtheorem{theorem}{Theorem}

\newtheorem{corollary}{Corollary}

\newcommand{\rnni}{\mathrm{RNNI}}
\newcommand{\findpath}{\textsc{FindPath}}

\newcommand{\rank}{\mathrm{rank}}
\newcommand{\ntime}{\mathrm{time}}
\newcommand{\nni}{\mathrm{NNI}}

\newcommand{\fp}{\mathrm{FP}}
\newcommand{\dtt}{\mathrm{DCT}}
\newcommand{\np}{\mathbf{NP}}

\renewcommand{\O}{\mathcal O}
\renewcommand{\epsilon}{\varepsilon}

\newcommand{\summary}[1]{} 

\DeclareMathOperator*{\argmax}{argmax}

\graphicspath{{figures/}}

\title[Geometry of ranked tree spaces]{The Geometry of the space of Discrete Coalescent Trees}
\date{\today}
\author{Lena Collienne\textsuperscript{1}}
\email{lcollienne@cs.otago.ac.nz}
\address{\textsuperscript{1}Department of Computer Science, University of Otago, New Zealand}
\author{Kieran Elmes\textsuperscript{1}}
\email{kelmes@cs.otago.ac.nz}
\author{Mareike Fischer\textsuperscript{2}}
\email{email@mareikefischer.de}
\address{\textsuperscript{2}Institute of Mathematics and Computer Science, University of Greifswald, Germany}
\author{David Bryant\textsuperscript{3}}
\email{david.bryant@otago.ac.nz}
\address{\textsuperscript{3}Department of Mathematics and Statistics, University of Otago, New Zealand}
\author{Alex Gavryushkin\textsuperscript{1, \Letter}}
\email{\textsuperscript{\Letter}alex@biods.org}
\thanks{}

\begin{document}

\begin{abstract}
Computational inference of dated evolutionary histories relies upon various hypotheses about RNA, DNA, and protein sequence mutation rates.
Using mutation rates to infer these dated histories is referred to as molecular clock assumption.
Coalescent theory is a popular class of evolutionary models that implements the molecular clock hypothesis to facilitate computational inference of dated phylogenies.
Cancer and virus evolution are two areas where these methods are particularly important.

Methodologically, phylogenetic inference methods require a tree space over which the inference is performed, and geometry of this space plays an important role in statistical and computational aspects of tree inference algorithms.
It has recently been shown that molecular clock, and hence coalescent, trees possess a unique geometry, different from that of classical phylogenetic tree spaces which do not model mutation rates.

Here we introduce and study a space of discrete coalescent trees, that is, we assume that time is discrete, which is inevitable in many computational formalisations.
We establish several geometrical properties of the space and show how these properties impact various algorithms used in phylogenetic analyses.
Our tree space is a discretisation of a known time tree space, called $t$-space, and hence our results can be used to approximate solutions to various open problems in $t$-space.
Our tree space is also a generalisation of another known trees space, called the ranked nearest neighbour interchange space, hence our advances in this paper imply new and generalise existing results about ranked trees.
\end{abstract}

\maketitle

\section{Introduction}

\summary{Molecular clock, divergence dating, and coalescent -- biological motivation}
A commonly used hypothesis in various applications in evolutionary biology is the molecular clock.
For example, a strict molecular clock is the assumption that the mutation rate of a gene is approximately constant over time.
After this phenomenon had first been observed by \textcite{zuckerkandl1965evolutionary}, the molecular clock became a popular hypothesis, and various relaxations were developed \autocite{Kumar2016-eu}.
A popular framework for reconstructing and analysing timed evolutionary (species) trees \autocite{Kingman1982-df} that uses the molecular clock assumption on gene trees is coalescent theory.
For example, coalescent is widely employed for inferring relationships of a sample of genes \autocite{Hudson1990-ki, Kuhner2009-jb}, or for analysing population dynamics \autocite{Kuhner1998-eh,Drummond2005-ak}.
A recent striking application of coalescent theory is cancer phylogenetics \autocite{Posada2020-aa, Ohtsuki2017-su}, where accurate estimates of divergence times are essential for targeted treatment strategies.
Under a coalescent model evolution is considered backwards in time, and two lineages coalesce after a waiting time, which is to be estimated.

\summary{Software needs to deal with clock trees, tree proposals}
In phylogenetic trees, which display evolutionary relationships, internal nodes can hence be equipped with times, when assuming a molecular clock.
Software packages for reconstructing those trees from data such as RNA, DNA, or protein sequences rely on a parameterisation of trees where internal nodes are equipped with times.
Popular tree inference software used for this purpose are based on Maximum Likelihood \autocite{Kozlov2019-cf, Nguyen2015-sp, Tamura2011-ky} or Bayesian methods \autocite{Bouckaert2014-ir,Suchard2018-tw, Ronquist2003-eq}.
They rely on tree search algorithms, where in every step a new tree is proposed and accepted if the proposed tree fulfils certain requirements.
For tree proposals under the molecular clock assumption a parameterisation of trees taking the times of internal nodes into account is required.
Furthermore, a similarity measure for these trees is necessary, to propose trees that are measurably similar to the running tree.

\summary{Known tree spaces -- BHV, $t$-space and $\tau$-space}
Tree spaces that take branch lengths of trees into account exist in the literature.
For example, the BHV-space \autocite{Billera2001-rj} models trees as points in a cubical complex.
However, this parameterisation is not suitable for coalescent trees because changing the times of an evolutionary event in the tree implies that all preceding events change their times as well.
Hence two trees can be close to each other in this space even though the timing of many internal nodes is different in the two trees.
Examples of more suitable tree spaces where internal nodes of trees are equipped with times are $t$-space and $\tau$-space \autocite{Gavryushkin2016-uu}.
It has been observed, however, that in the $\tau$-space, similarly to the BHV-space, shortest paths between trees often contain the star tree \autocite{Gavryushkin2016-uu}, a property that can be problematic in applications.
Although the $t$-space is free from these properties, no algorithm for computing distances or shortest paths between trees in this space is known yet, so applications are limited.

\summary{Why we want to investigate geometrical properties of $\dtt_m$ and $\rnni$}
Enabling statistical analysis over the space of phylogenetic trees was an important motivation for \textcite{Billera2001-rj} to introduce the BHV-space and study its geometric properties.
Tree space geometry has also played an important role in studies of rogue taxa in a tree \autocite{Cueto2011-bh} and also summary trees \autocite{Miller2015-rk}.
Here, driven by the same motivation, we propose to study coalescent trees.

\summary{Structure of the paper.}
In this paper we introduce the space $\dtt_m$ of discrete coalescent trees, where internal nodes are assigned unique discrete times.
This tree space is a discrete version of the $t$-space.
$\dtt_m$ is also a generalisation of the ranked nearest neighbour interchange ($\rnni$) space \autocite{Collienne2020-iu}.
Here we show that the space $\dtt_m$ as well as $\rnni$ have the desired properties mentioned above, including efficiently computable shortest paths that preserve biological information shared between trees.
After introducing notations used throughout this paper (\autoref{section:technical_introduction}), we discuss how the algorithm $\findpath$ \autocite{Collienne2020-iu} can be generalise from $\rnni$ to be applied to discrete coalescent trees, computing shortest paths in polynomial time (\autoref{section:fp_dtt}).
We then analyse some geometrical properties of both tree spaces $\dtt_m$ and $\rnni$ (\autoref{section:geometry}) -- first, we discuss the cluster property in \autoref{section:cluster_property} and then consider a subset of trees (caterpillar trees) for which we are able to compute $\rnni$ distances more efficiently than with $\findpath$ (\autoref{section:caterpillar_convex}).
Following that, we establish the diameter of $\dtt_m$ and $\rnni$ and briefly discuss the radius for each space.
We finish this paper with a section providing a connection between the $\rnni$ space and partition lattices, and propose directions for further research (\autoref{section:open_problems}).

\section{Technical Introduction}
\label{section:technical_introduction}

\summary{Introducing discrete coalescent trees and ranked trees}
A \emph{rooted binary phylogenetic tree} is a binary tree with $n$ leaves uniquely labelled by elements of a set $\{a_1, \ldots, a_n\}$.
The main object of study in this paper are \emph{discrete coalescent trees}, binary rooted phylogenetic trees with a positive integer-valued \emph{time} assigned to each node.
More specifically, all $n$ leaves $a_1, \ldots, a_n$ are assigned time $0$, and every internal node is assigned a unique time less or equal to an integer $m$, such that it always has time greater than its children.
Note that this implies $m \geq n-1$.
We denote the time of an internal node $v$ by $\ntime(v)$.
If not stated otherwise, we refer to discrete coalescent trees simply as \emph{trees}.
We furthermore call two trees (not necessarily binary) \emph{identical} if there is a graph isomorphism between them preserving leaf labels and times.

\begin{figure}[ht]
	\includegraphics[width=0.25\textwidth]{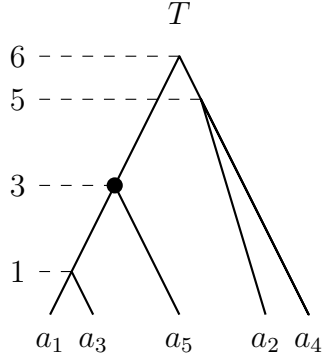}
	\caption{Discrete coalescent tree with $n = 5$ leaves and root height $m = 6$.
	The highlighted node with time three can be referred to as $(a_5)_T$, $(\{a_1,a_5\})_T$, and the cluster induced by this node is $(T)_3$.}
	\label{fig:dtt_tree}
\end{figure}

As a special case of discrete coalescent trees we consider \emph{ranked trees} with root time $n-1$.
In these trees internal nodes have distinct times ranging from $1$ to $n-1$.
This definition of ranked trees coincides with the one of \textcite{Collienne2020-iu}.
In the case of ranked trees we say \emph{rank} of a node $v$ to mean its time ($\rank(v) = \ntime(v)$) to be consistent with notations used in \autocite{Collienne2020-iu}.
There are $\frac{(n-1)!n!}{2^{n-1}}$ ranked trees \autocite{Semple2003-nj}.
Every ranked tree gives ${m \choose n-1}$ discrete coalescent trees, as every $(n-1)$-element subset of $\{1, \ldots, m\}$ can be the set of times assigned to the internal nodes of a ranked tree.
Hence there are, contrary to the claim in \autocite{Gavryushkin2018-ol}, $\frac{(n-1)!n!}{2^{n-1}} {m \choose n-1}$ discrete coalescent trees.

Every internal node $v$ of a tree $T$ can be referred to by the set $C$ of leaves that are descending from this node.
We call such a set $C$ \emph{cluster} and say that the cluster $C$ is \emph{induced} by $v$.
A list of clusters $[C_1, \ldots, C_{n-1}]$ determines at most one ranked tree \autocite{Collienne2020-iu}, where cluster $C_i$ is induced by the internal node with rank $i$ for $i \in \{1, \ldots, n-1\}$.
For discrete coalescent trees however, times of nodes also need to be provided to uniquely identify a tree.
For a subset $S \subseteq \{a_1, \ldots, a_n\}$ we call the internal node of a tree $T$ with lowest time among those ancestral to all elements of $S$ the \emph{most recent common ancestor} of $S$ and denote it by $(S)_T$.
We furthermore denote the parent of a leaf $a_i$ in $T$ by $(a_i)_T$, and the cluster induced by the node with time $i$ in $T$ by $(T)_i$.
The node highlighted in \autoref{fig:dtt_tree} for example can be referred to as $(a_5)_T$, the parent of $a_5$, or $(\{a_1, a_5\})_T$, the most common ancestor of $\{a_1, a_5\}$, or $(T)_3$, the node with time three in $T$.
Note that we will simply write $\rank(a_i)_T$ or $\ntime(a_i)_T$ to mean $\rank((a_i)_T)$ or $\ntime((a_i)_T)$, respectively.
Although differing from traditional notations, our notation with brackets referring to internal nodes is intuitive, shortens nested formulas, and is consistent with notations used in \autocite{Collienne2020-iu}.
A type of trees that will be of importance throughout the whole paper are \emph{caterpillar trees}, which are trees where every internal nodes has at least one child that is a leaf.

\begin{figure}[ht]
	\includegraphics[width=0.85\textwidth]{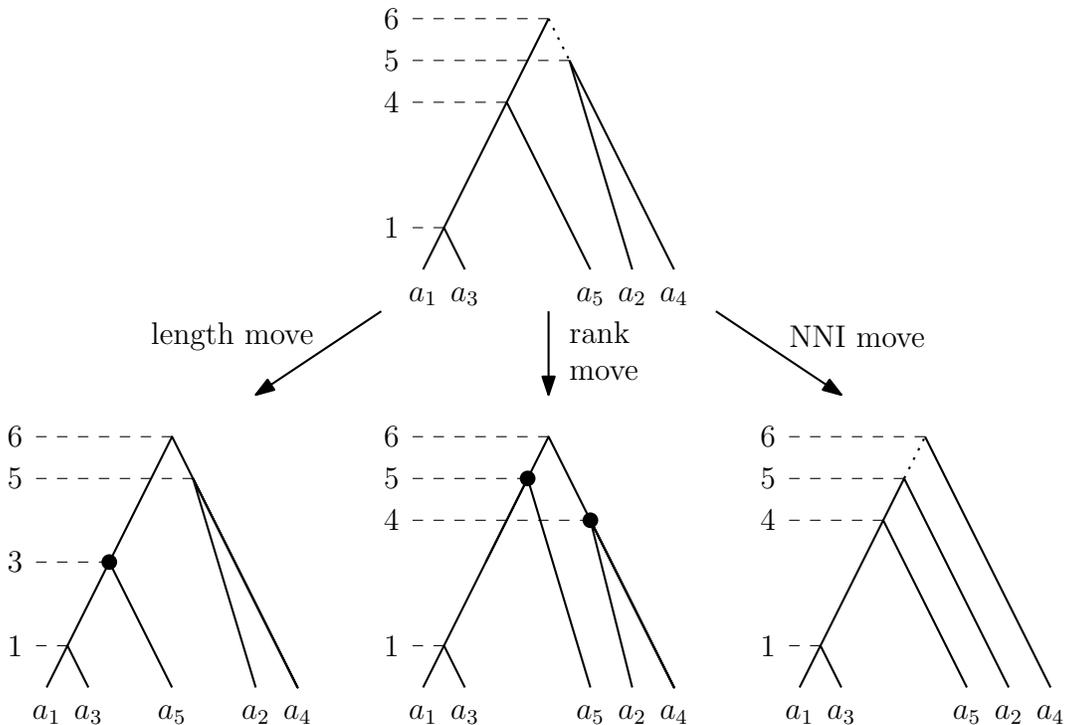}
	\caption{The three possible moves on a discrete coalescent tree: a length move changing the time of the highlighted node on the left, a rank move swapping the ranks of the highlighted nodes 	in the middle and an $\nni$ move on the dotted edge on the right.}
	\label{fig:dtt}
\end{figure}

\summary{Defining the tree space $\dtt_m$ and $\rnni = \dtt_{n-1}$}
We are now ready to introduce the central object of study of this paper, the graph (or space) of discrete coalescent trees.
This graph is called $\dtt_m$ for a fixed positive integer $m$.
The vertex set of $\dtt_m$ is the set of trees with root time less or equal to $m$.
Note that a second parameter of $\dtt_m$ is the number of leaves $n$ of the trees in the graph, which we assume to be fixed throughout this paper.
Trees $T$ and $R$ are connected by an edge ($T$ and $R$ are \emph{neighbours}) in this graph if performing one of the following (reversible) operations on $T$ results in $R$ (\autoref{fig:dtt}):
\begin{enumerate}
	\item An \emph{$\nni$ move} connects trees $T$ and $R$ if there is an edge $e$ in $T$ and an edge $f$ in $R$, both of length one, such that shrinking $e$ and $f$ to nodes results in identical trees.
	\item A \emph{rank move} on $T$ exchanges the times of two internal nodes with time difference one.
	\item A \emph{length move} on $T$ changes the time of an internal node by one.
\end{enumerate}
A length move can only change the time of a node to become $t$ if there is no node with time $t$ already.
Furthermore, the time of the root of a tree in $\dtt_m$ cannot be changed by a length move to become greater than $m$ in $\dtt_m$.
Note that our definition of $\dtt_m$ differs from the definition of the space on discrete time-trees of \textcite{Gavryushkin2018-ol}.
In contrast to their definition, length moves in $\dtt_m$ do not change the height of a tree, unless it is performed on the root, which makes our definition appropriate for coalescent trees.

The definition of $\dtt_m$ leads to a natural definition of the \emph{distance} between two trees $T$ and $R$ in this graph as the length of a shortest paths between these trees, denoted by $d(T,R)$.
We also consider the ranked nearest neighbour interchange ($\rnni$) graph of \textcite{Collienne2020-iu}, which is the graph $\dtt_m$ for $m=n-1$, and hence a graph of ranked trees.
In this graph length moves are not possible, so we use the notion $\rnni$ \emph{move} to mean either a rank move or an $\nni$ move in order to distinguish these moves from length moves.

\section{Computing Shortest Paths in $\dtt_m$}
\label{section:fp_dtt}

\summary{Introduce how we can use $\findpath$ to compute $\dtt_m$ distances}
Shortest paths, and therefore distances, between trees in $\rnni$ can be computed with the algorithm $\findpath$, which was introduced by \textcite{Collienne2020-iu} and has running time quadratic in the number of leaves $n$.
As $\rnni$ is a special case of $\dtt_m$ for $m = n-1$, the question arises whether a modification of this algorithm can also be used to compute shortest paths in $\dtt_m$.
In this section we present a generalisation of $\findpath$ that computes distances between trees in $\dtt_m$.
Before introducing the version of $\findpath$ for $\dtt_m$, we introduce a way to convert trees in $\dtt_m$ on $n$ leaves into ranked trees on $m+2$ leaves, such that the $\rnni$ distance between those ranked trees equals their distance in $\dtt_m$ (\autoref{thm:dtt_findpath}).

\begin{figure}[ht]
	\includegraphics[width=0.75\textwidth]{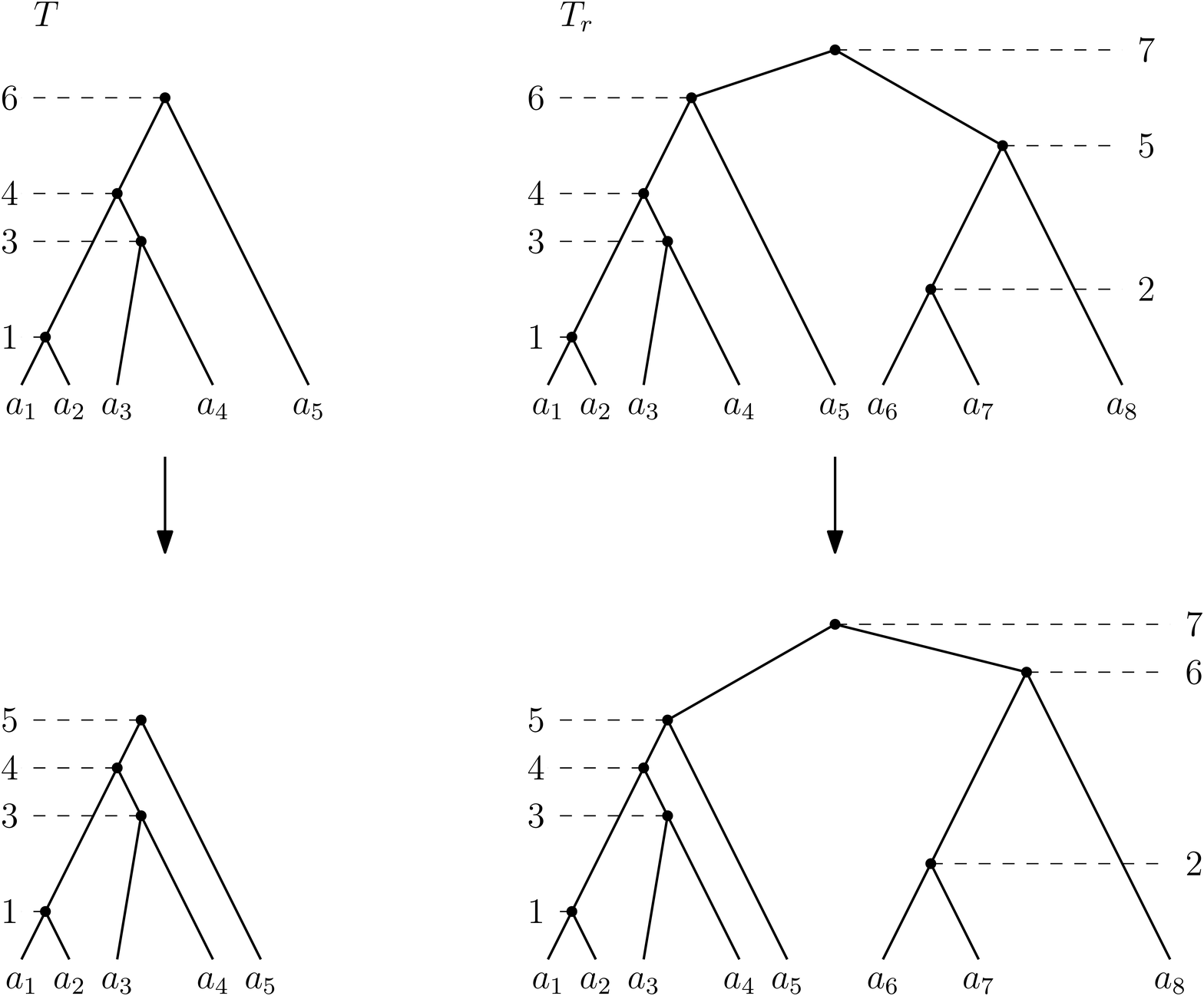}
	\caption{Extending a tree $T$ on $n$ leaves in $\dtt_6$ (top left) to a ranked tree with $m+2=8$ leaves (top right) by adding a caterpillar subtree with three leaves.
	The trees on the bottom result from $T$ and $T_r$ by performing a length move (left) or rank move (right), respectively.}
	\label{fig:dtt_to_ranked_tree}
\end{figure}

\summary{How to add leaves to a $\dtt_m$ tree to transform it into a ranked tree}
A tree $T$ in $\dtt_m$ on $n$ leaves can be converted into a ranked tree in $\rnni$ with $m+2$ leaves in the following way (\autoref{alg:ranked_tree}).
First add a new root with time $m + 1$ that becomes parent of the root of $T$.
The other child of this new root becomes the root of a caterpillar tree ${T_r}^c$ on leaf set $\{a_{n+1}, a_{n+2}, \ldots, a_{m+2}\}$, such that $\ntime(a_{n+1})_{{T_r}^c} = \ntime(a_{n+2})_{{T_r}^c} < \ntime(a_{n+3})_{{T_r}^c} < \ldots < \ntime(a_{m+2})_{{T_r}^c} < m+1$.
An example of this extension of a tree $T$ to a ranked tree $T_r$ is depicted in \autoref{fig:dtt_to_ranked_tree}.

Throughout this paper we denote this extended ranked version of a tree $T$ by $T_r$.
Moreover, we denote the subtree of $T_r$ that is identical to $T$ by $T_r^d$ ($d$ for discrete coalescent tree) and the caterpillar subtree on leaf set $\{a_{n+1}, \ldots, a_{m+2}\}$ by $T_r^c$.

\begin{algorithm}[ht]
	\caption{RankedTree($T$, $m$)}
	\label{alg:ranked_tree}
	\begin{algorithmic}[1]
		\STATE $S:= \{1 \leq i \leq m \ |\  \text{ no internal node in } T \text{ has time } i\}$
		\STATE $[i_1, \ldots, i_{m-n+1}] = sort(S)$
		\STATE $T_r^d =$ copy of $T$
		\STATE $T_r^c =$ tree consisting of just one internal node $v_1$ with rank $i_1$ and children $a_{n+1}, a_{n+2}$
		\FOR {$k = 2, \dots, m-n+1$}
			\STATE Add internal node $v_k$ with with time $i_k$ and children $v_{k-1}$ and $a_{n+1+k}$ to $T_r^c$
		\STATE $T_r = $ tree with root with time $m+1$ and children of root are roots of $T_r^d$ and $T_r^c$.
		\ENDFOR
		\RETURN $T_r$
	\end{algorithmic}
\end{algorithm}

\summary{Moves on the extended ranked versions of trees -- $\rnni$ vs length moves}
In the following we distinguish two different types of rank moves.
Rank moves between one node of $T_r^c$ and one node of $T_r^d$ induce length moves on the subtree $T_r^d$ in $\dtt_m$ (\autoref{fig:dtt_to_ranked_tree}).
Therefore, we will refer to such rank moves as \emph{rank moves corresponding to length moves}.
All remaining rank moves will still be called rank moves.
Note that the correspondence of rank moves between $T_r^c$ and $T_r^d$ to length moves in $T$ shows that any path between $T$ and $R$ in $\dtt_m$ can be interpreted as a path between $T_r$ and $R_r$ in $\rnni$.

\summary{How to compute shortest $\dtt_m$-paths between trees with $\findpath$}
After extending both trees $T$ and $R$ in $\dtt_m$ to ranked trees $T_r$ and $R_r$ on $m+2$ leaves, respectively, we can compute shortest paths between $T_r$ and $R_r$ in $\rnni$, using $\findpath$.
A path computed by $\findpath$ preserves clusters \autocite{Collienne2020-iu}, hence there are no $\nni$ moves in the newly added caterpillar subtree on the leaf set $\{a_{n+1}, \ldots, a_{m+2}\}$ on such a path.
The only moves involving internal nodes of this caterpillar subtree are rank moves corresponding to length moves, as described above.
Hence the path $\fp(T_r,R_r)$ provides a path between $T$ and $R$ in $\dtt_m$, when only considering the subtrees induced by $\{a_1, \ldots, a_n\}$ in all trees on $\fp(T_r, R_r)$, interpreting some rank moves between $T_r$ and $R_r$ as length moves.
We denote this $\dtt_m$ path, which results from $\fp(T_r, R_r)$, by $\fp(T,R)$.
In \autoref{thm:dtt_findpath} we establish that $\fp(T,R)$ is indeed a shortest path in $\dtt_m$.
Note that for any given pair of trees $T$ and $R$, we always assume $m$ to be the maximum root time of these trees and consider a shortest path between them in $\dtt_m$.

\begin{theorem}
	The path $\fp(T,R)$ between two discrete coalescent trees $T$ and $R$ is a shortest path in $\dtt_m$, where $m$ is the maximum root time of $T$ and $R$.
	\label{thm:dtt_findpath}
\end{theorem}

\begin{proof}
	Let $T$ and $R$ be discrete coalescent trees and $T_r$ and $R_r$ their extended ranked versions computed with \autoref{alg:ranked_tree}, respectively.
	Any path in $\dtt_m$ from $T$ to $R$ gives a path of equal length between $T_r$ and $R_r$ in the $\rnni$ space on $m+2$ leaves.
	This is due to the fact that the only moves needed in the subtree $T_r^c$ to transform it to $R_r^c$ are rank moves corresponding to length moves, and no other $\rnni$ moves.
	If there was a path between $T$ and $R$ shorter than $\fp(T,R)$, the corresponding path between $T_r$ and $R_r$ in $\rnni$ would be shorter than the one computed by $\findpath$ in this space.
	Since this contradicts the fact that $\findpath$ computes shortest paths in $\rnni$ \autocite[Theorem 1]{Collienne2020-iu}, it follows that $\fp(T,R)$ is a shortest path in $\dtt_m$.
\end{proof}

\summary{Running time of $\findpath$ + we don't need to add subtree in practice}
\autoref{thm:dtt_findpath} shows that $\findpath$ computes a shortest path between two trees in $\dtt_m$ in polynomial time, more specifically in $\O(mn)$.
More details on the running time are discussed in \autoref{section:diameter} following \autoref{thm:dtt_diameter}.
It is not even necessary to convert a given pair of discrete coalescent trees to ranked trees to apply $\findpath$ to them.
Instead, we modify $\findpath$ for trees in $\dtt_m$ (\autoref{alg:fp_dtt}).
Iterations of $\findpath$ that consider clusters in the added caterpillar trees are replaced by length moves increasing the time of internal nodes as described in the \textbf{for} loop in Line~\ref{line:length_move} of \autoref{alg:fp_dtt}.
The benefit of this modified version of the algorithm, compared to using $\findpath$ on the extended ranked versions of the trees, is a reduced use of memory, which is especially of practical relevance for $m \gg n$, which is typical in applications.

Note that we do not need the parameter $m$ in practice, as the distance between any two trees in $\dtt_{m'}$ is the same as their distance in $\dtt_m$ for any $m > m'$.
Therefore, if the distance between two trees is to be computed, we can simply choose $m$ to be the maximum root height of the given trees and compute their distance in $\dtt_m$.

\begin{algorithm}[h]
	\caption{$\findpath$($T,R$)}
	\begin{algorithmic}[1]
		\label{alg:fp_dtt}
		\STATE $T_1 := T$, $p := [T_1]$
		\FOR {$k = 1, \dots, m$}
			\IF {$R$ has a node with time $k$}
			\STATE $C:=(R)_k$
			\WHILE {$\ntime((C)_{T_1})>k$}
					\STATE $T_2$ is $T_1$ with the time of $(C)_{T_1}$ decreased by an $\rnni$ move
				\STATE $T_1 = T_2$
				\STATE $p = p+T_1$
			\ENDWHILE
			\ELSIF {$T$ has a node with time $k$}
				\STATE $i := \min\{l \ |\  l>k \text{ and no node in } T_1 \text{ has time }l\}$
				\FOR {$j = i-1, \dots, k$}
					\label{line:length_move}
					\STATE $T_2$ is $T_1$ where the time of $(T_1)_j$ is increased by one (length move)
					\STATE $T_1 = T_2$
					\STATE $p = p+T_1$
				\ENDFOR
			\ENDIF
		\ENDFOR
		\RETURN $p$
	\end{algorithmic}
\end{algorithm}

\section{Geometrical Properties of $\dtt_m$}
\label{section:geometry}

\subsection{Cluster Property}
\label{section:cluster_property}
\summary{Definition of Cluster Property and why it is relevant (a bit of bio).}
A tree space has the \emph{cluster property}, if all trees on every shortest path between two trees sharing a cluster $C$ also contain $C$.
This is a desirable property in evolutionary biology applications as trees sharing a cluster or subtree are expected to be closer to each other than to a tree not sharing a cluster with them.
This property is also desirable in centroid-based tree sample summary methods.
For a given sample of trees containing a common subtree, it is expected that their summary tree also contains this subtree.
It is therefore desirable to have a tree space that has the cluster property.

\summary{Cluster property in $\nni$ and its connection to the complexity result.}
A mathematical motivation for investigating the cluster property in $\rnni$ is its importance in a similar tree space, the nearest neighbour interchange graph ($\nni$).
In the $\nni$ graph, trees have no times and $\nni$ moves are allowed on every edge.
Computing distances in $\nni$ is $\np$-hard \autocite{Dasgupta2000-xa}, and the proof relies on the fact that this tree space does not have the cluster property \autocite{Li1996-zw}.
In the $\rnni$ graph, however, distances can be computed in polynomial time using $\findpath$ \autocite{Collienne2020-iu}, which preserves common clusters.
The question whether $\rnni$ has the cluster property is hence natural, and will be settled in \autoref{thm:cluster_property_rnni}.

\summary{$\rnni$ has the cluster property.}
\begin{theorem}
	The $\rnni$ graph has the cluster property.
	\label{thm:cluster_property_rnni}
\end{theorem}

\begin{proof}
	We assume to the contrary that there are two ranked trees $T$ and $R$ sharing a cluster $C$ and a shortest path $p$ between these trees where $C$ is not present in every tree.
	We furthermore assume that there is no pair of trees with a shorter path not containing a shared cluster and distance less than $d(T,R)$, meaning that $T$ and $R$ give a minimum counterexample.
	Because of this minimality assumption on the length of $p$, the first tree $T'$ following $T$ on $p$ does not contain $C$.
	Since $C$ must be the only cluster changed by the $\nni$ move between $T$ and $T'$, all nodes with rank below $(C)_T$ induce the same clusters in $T$ and $T'$ (\autoref{fig:nni_neighbours_clusters}).
	We now compare distances $d(T,R)$ and $d(T',R)$ by using properties of $\findpath$.

	\begin{figure}[ht]
		\includegraphics[width=0.33\textwidth]{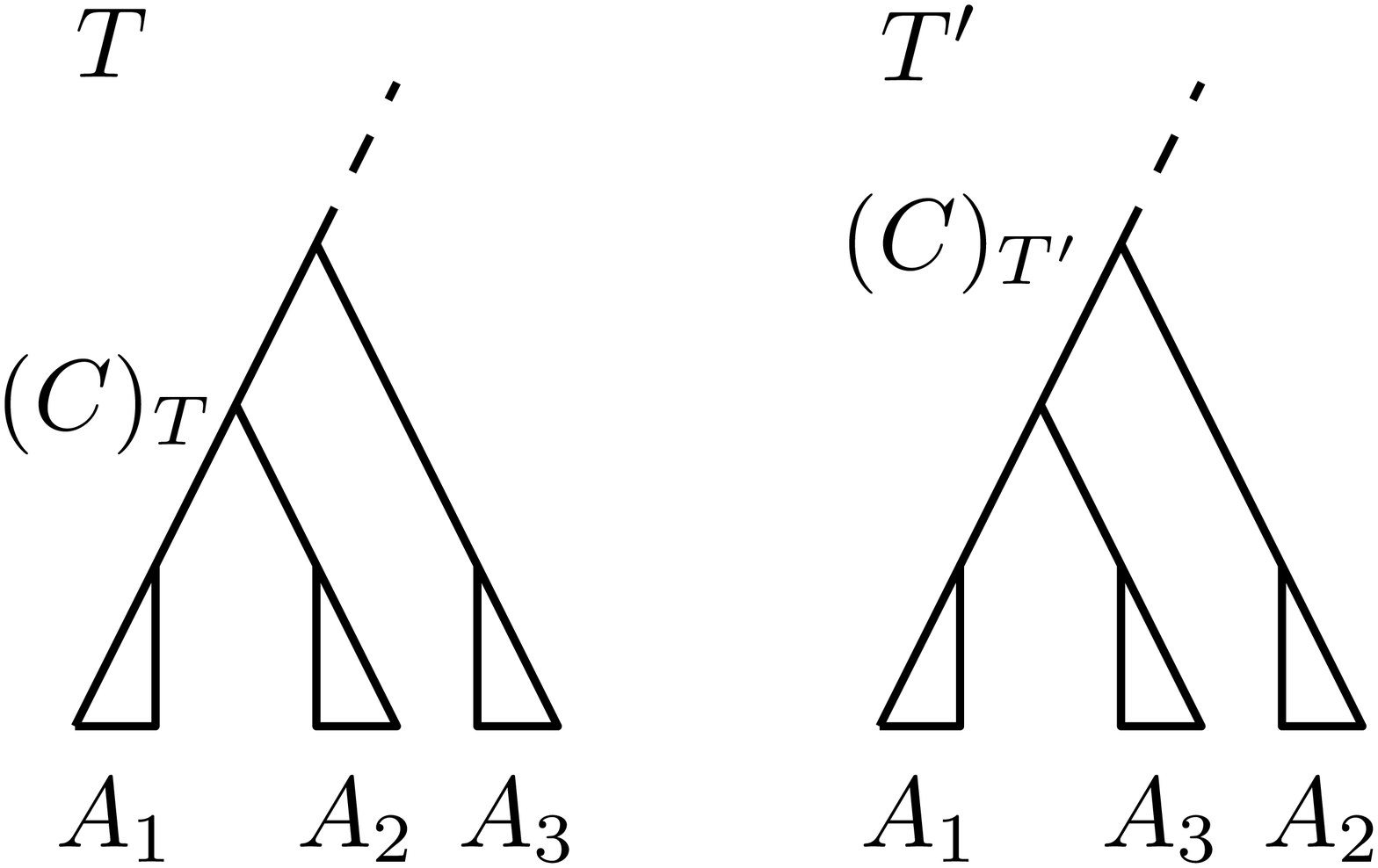}
		\caption{Trees $T$ and $\nni$ neighbour $T'$, such that the cluster $C = A_1 \cup A_2$ is not present in $T'$, but in $T$.}
		\label{fig:nni_neighbours_clusters}
	\end{figure}

	First we compare $\fp(R,T)$ and $\fp(R,T')$.
	All trees on these two paths coincide up to iteration $i = \rank((C)_T)$, in which the cluster considered on $\fp(R,T)$ is $C$.
	Let $R'$ denote the tree at this point of the path, meaning that $\fp(R,T)$ and $\fp(R,T')$ coincide up to this tree $R'$.
	It follows $d(T,R) = d(R,R') + d(R', T)$ and $d(T',R) = d(R,R') + d(R', T')$.

	Now consider $\fp(T', R')$ to evaluate $d(R', T')$.
	As $\findpath$ preserves clusters, $C$ is present in every tree on $\fp(T,R)$ up to and including $R'$.
	The first iteration of $\findpath$ applied to the pair of trees $(T',R')$ hence considers the cluster $C$, as all cluster induced by nodes below $(C)_{T'}$ coincide in $R'$ and $T'$.
	To construct the cluster $C$ in $T'$, there is just one $\nni$ move needed, which results in the tree $T$, as $T$ and $T'$ are $\nni$ neighbours such that $T$ contains $C$ and $T'$ does not (\autoref{fig:nni_neighbours_clusters}).
	We can therefore conclude that $d(T,R) = d(T',R) - 1$, which contradicts the assumption that $T'$ is the first tree on a shortest path from $T$ to $R$.
	There is hence no shortest path between $T$ and $R$ that does not preserve $C$, which proves the cluster property for $\rnni$.
\end{proof}

The fact that the slightly modified version of $\findpath$ computes shortest paths in $\dtt_m$ already suggests that shortest paths in $\rnni$ and $\dtt_m$ have similar properties.
Indeed, the cluster property in $\dtt_m$ follows from \autoref{thm:cluster_property_rnni}.

\begin{corollary}
	The graph $\dtt_m$ has the cluster property.
\end{corollary}

\begin{proof}
	Assume that there is a shortest path between two trees $T$ and $R$ in $\dtt_m$ that does not preserve a common cluster.
	This path corresponds to a path between $T_r$ and $R_r$, the extended ranked versions of $T$ and $R$ in $\rnni$, as already discussed in \autoref{thm:dtt_findpath}.
	Since this path has the same length as the one between $T_r$ and $R_r$, it is be a shortest path in $\rnni$ as well, which leads to a contradiction to \autoref{thm:cluster_property_rnni}.
\end{proof}

\subsection{Caterpillar Trees}
\label{section:caterpillar_convex}

\summary{Defining Caterpillar trees. Why are they interesting?}
In this subsection we focus on the set of caterpillar trees and establish some properties of shortest paths between those trees in both $\rnni$ and $\dtt_m$.
In \autoref{thm:caterpillar_convex_dtt} we will see that, in both $\dtt_m$ and $\rnni$, any two caterpillar trees are connected by a shortest path consisting only of caterpillar trees.
We say that a set of trees is \emph{convex} in a tree space, if there is a shortest path between any two trees in this set that stays within the set.
The set of caterpillar trees is hence convex in $\rnni$.
The $\nni$ space of unranked trees however does not have this property \autocite{Gavryushkin2018-ol}.
Based on the convexity of the set of caterpillar trees in $\rnni$ we introduce a way to compute distances between caterpillar trees in this space in time $\O(n \sqrt{\log n})$ in \autoref{cor:caterpillar_distance_rnni_nlogn}, and hence with better worst-case time complexity than $\findpath$.
Whether this complexity can be achieved in $\dtt_m$ is an open question.

\begin{theorem}
	The set of caterpillar trees is convex in $\dtt_m$.
	\label{thm:caterpillar_convex_dtt}
\end{theorem}

\begin{proof}
	Let $T$ and $R$ be two caterpillar trees in $\dtt_m$.
	We prove the theorem by showing that there is a caterpillar tree $T'$ that is a neighbour of $T$ and closer to $R$ than $T$.
	The existence of a shortest path consisting only of caterpillar trees between $T$ and $R$ follows inductively.
	Throughout this proof we consider the extended ranked versions $T_r$ and $R_r$ of $T$ and $R$.

	Let $a_k := \argmax_{a_1, \ldots, a_n}\{\rank(a_i)_{R_r} \ |\  \rank(a_i)_{R_r} \neq \rank(a_i)_{T_r}\}$ be the leaf with parent with maximum rank in $R_r$ among those whose parents do not have equal rank in $T_r$ and $R_r$.
	Let furthermore $a_j \in \{a_1, \ldots, a_{m+2}\}$ be a leaf with $\rank(a_j)_{T_r} = \rank(a_k)_{T_r} + 1$.
	We define $T'_r$ to be the caterpillar tree resulting from $T_r$ by an $\nni$ move or rank move exchanging the ranks of $(a_k)_{T_r}$ and $(a_j)_{T_r}$.
	An $\nni$ move is necessary if these two nodes are connected by an edge, otherwise a rank move corresponding to a length move is performed on $T_r$ to obtain $T'_r$ (\autoref{fig:caterpillar_convex}).
	In both cases ${T'_r}^d$ is a caterpillar tree.
	We will use properties of shortest paths computed by $\findpath$ to show that $|\fp(R_r,T'_r)| = |\fp(R_r,T_r)| - 1$.

	\begin{figure}[ht]
		\includegraphics[width=1\textwidth]{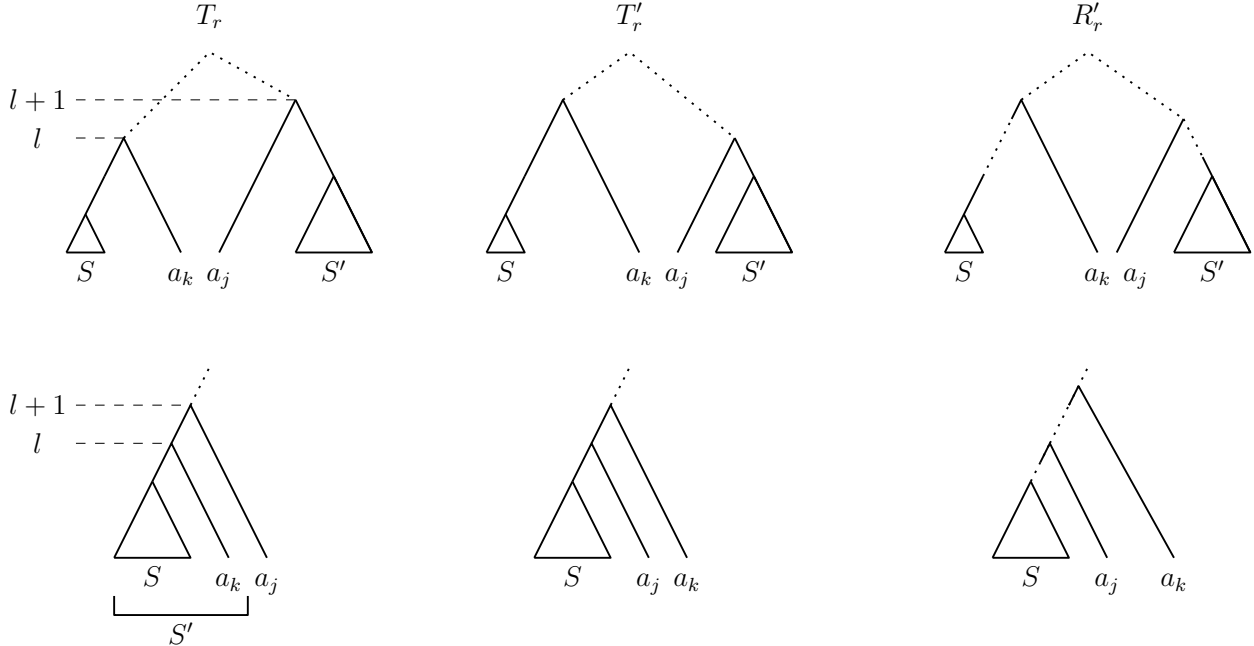}
		\caption{The two possible versions of trees $T_r$ (left), $T'_r$ (middle), and $R'_r$ as described in the proof of \autoref{thm:caterpillar_convex_dtt}.
		Between $T_r$ and $T'_r$ only the ranks of the parents of $a_j$ and $a_k$ are exchanged, the rest of the trees coincide.
		At the bottom the case that $(a_j)_T$ is parent of $(a_k)_T$ and $S' = S \cup \{a_k\}$ is displayed.
		$S'$ is a cluster in all three trees at the bottom.
		At the top $(a_j)_T$ and $(a_k)_T$ are in the two different subtrees $T_r^d$ and $T_r^c$ (the same in $T'_r$ and $R'_r$), which is also true for the disjoint sets $S$ and $S'$, which are present as clusters in all three trees.
		Dotted lines represent remaining parts of trees, which are equal in $T_r$ and $T'_r$, but different to $R'_r$.
		Note that the rank difference of $(a_k)_{R'_r}$ and $(a_j)_{R'_r}$ does not need to be one, which it is in $T_r$ and $T'_r$.
		\label{fig:caterpillar_convex}}
	\end{figure}

	Since all clusters of $T_r$ and $T'_r$ induced by nodes of rank less than $\rank(a_k)_{T_r}$ coincide, the paths $\fp(R_r,T_r)$ and $\fp(R_r,T'_r)$ coincide up to a ranked tree $R'_r$, which contains all these clusters.
	We now compare the lengths of $\fp(R'_r,T_r)$ and $\fp(R'_r,T'_r)$.
	We note at first that $\rank(a_j)_{R_r} < \rank(a_k)_{R_r}$.
	If it otherwise was $\rank(a_k)_{R_r} \leq \rank(a_j)_{R_r}$, it would follow $\rank(a_j)_{R_r} = \rank(a_j)_{T_r}$, by the definition of $a_k$, and therefore  $\rank(a_k)_{R_r} \leq \rank(a_j)_{R_r} = \rank(a_j)_{T_r} = \rank(a_k)_{T_r} + 1$.
	$\rank(a_k)_{R_r} \leq \rank(a_k)_{T_r} + 1$ however contradicts the definition of $a_k$, hence $\rank(a_j)_{R_r} < \rank(a_k)_{R_r}$.
	It follows $\rank(a_j)_{R'_r} < \rank(a_k)_{R'_r}$, as $a_j$ and $a_k$ are not in any of the clusters considered by $\findpath$ before $R'_r$, which means that their parents do not exchange ranks before $R'_r$.

	By our assumptions on $T_r$, the cluster considered on $\fp(R_r,T_r)$ in iteration $l = \rank(a_k)_{T_r}$, which is the iteration following $R'_r$, is $S \cup \{a_k\}$, where $S$ is a cluster that is present in all three trees $T_r, T'_r,$ and $R_r$.
	In the following iteration $l+1 = \rank(a_j)_{T_r}$, $S' \cup \{a_j\}$ is considered for a cluster $S'$, where $S$ either equals $S \cup \{a_k\}$, if $T_r$ and $T'_r$ are connected by an $\nni$ move (bottom of \autoref{fig:caterpillar_convex}), or is a cluster present in ${T_r}^c$, ${T'_r}^c$, and ${R'_r}^c$, if $T_r$ and $T'_r$ are connected by a rank move (top of \autoref{fig:caterpillar_convex}).
	Decreasing the rank of $(S \cup \{a_k\})_{R'_r}$ takes $\rank(S \cup \{a_k\})_{R'_r} - l$ $\rnni$ moves.
	Because the rank of $(S \cup \{a_j\})_{R'_r}$ increases by one when the parents of $a_k$ and $a_j$ swap ranks in this iteration, the following iteration for $S' \cup \{a_j\}$ needs $\rank(S' \cup \{a_j\})_{R'_r} + 1 - (l+1)$ $\rnni$ moves.
	On $\fp(R_r,T'_r)$ however, first $\rank(S' \cup \{a_j\})_{R'_r} - l$ $\rnni$ moves decrease the rank of $(S' \cup \{a_j\})_{R'}$, and then $\rank(S \cup \{a_k\})_{R'_r} - (l+1)$ are needed for $S \cup \{a_k\}$.
	In total, these two iterations combined result in at least one extra move on $\fp(R_r, T_r)$ comparing to $\fp(R_r, T'_r)$.

	The only difference in the trees after iteration $l+1$ on the two different paths is the order of ranks of the parents of $a_j$ and $a_k$.
	Since the rest of $T_r$ and $T'_r$ coincide, the remaining parts of $\fp(R_r, T_r)$ and $\fp(R_r, T'_r)$ consist of the same moves.
	With our previous observation we can conclude $d(R_r,T_r) = d(R_r,T'_r) + 1$, and hence $T'_r$ is on a shortest path from $T_r$ to $R_r$.
\end{proof}

Note that it follows that the  set of caterpillar trees is convex in $\rnni$.
This convexity property implies that the distance between caterpillar trees can be computed more efficiently than by $\findpath$.
We prove this in the rest of this section.
To do so, we first establish that the problem of computing a shortest path consisting only of caterpillar trees can be interpreted in a few different ways.

One problem analogous to the shortest path problem for caterpillar trees in $\rnni$ is the \emph{Token Swapping Problem} \autocite{Kawahara2017-ey} on a special class of graphs, so-called lollipop graphs.
An instance of the token swapping problem is a simple graph where every vertex is assigned a token.
Two tokens are allowed to swap positions if they are on vertices that are connected by an edge.
Each token is assigned a unique goal vertex, and the aim is to find the minimum number of token swaps for all tokens to reach their goal vertex.

\begin{figure}[ht]
	\includegraphics[width=0.45\textwidth]{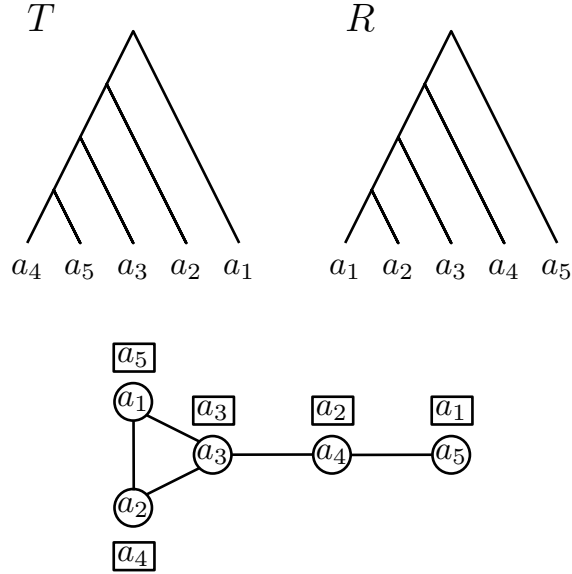}
	\caption{Two caterpillar trees $T$ and $R$ and the corresponding instance of the token swapping problem.
	Vertex labels are in circles and token goal vertices in rectangles.}
	\label{fig:tsp_caterpillar}
\end{figure}

The problems of computing distances between caterpillar trees can be seen as an instance of the token swapping problem on lollipop graphs.
A lollipop graph is a graph consisting of a complete graph that is connected to a path by one edge.
An instance of the token swapping problem that corresponding to the distance problem for caterpillar trees is described in the following.
An example is illustrated in \autoref{fig:tsp_caterpillar}.
Let $T$ and $R$ be caterpillar trees with
\[\rank(a_1)_R = \rank(a_2)_R < \rank(a_3)_R < \ldots < \rank(a_n)_R \text{ and}\]
\[\rank(b_1)_T = \rank(b_2)_T < \rank(b_3)_T < \ldots < \rank(b_n)_T\]
such that $[b_1, \ldots, b_n]$ is a permutation of $[a_1, \ldots, a_n]$.
The corresponding instance of the token swapping problem consists of a lollipop graph consisting of a complete graph on three leaves, connected to a path of length $n-3$ by an edge.
The vertex in the complete graph incident to the edge connecting complete graph and path is labelled by $a_3$, the other ones in the complete graph are labelled by $a_1$ and $a_2$.
The vertices on the paths are then labelled inductively, starting at the neighbour of $a_3$, such that the neighbour of the last already labelled node with label $a_{i-1}$ is labelled by $a_i$.
The token on vertex $a_i$ has $b_i$ as goal vertex.
Since the only moves between two caterpillar trees in $\rnni$ are $\nni$ moves, which simply swap two leaves, they correspond to swapping two tokens in the above described instance of the token swapping problem.

Therefore, the algorithm described by \textcite{Kawahara2017-ey} to solve the token swapping problem on lollipop graphs can be used for computing distances between caterpillar trees.
It however has worst-case time complexity $\O(n^2)$, the same as $\findpath$.

In the following we present an algorithm for computing distances between caterpillar trees with better worst-case time complexity, $\O(n \sqrt{\log n})$, for $\rnni$ (\autoref{cor:caterpillar_distance_rnni_nlogn}).
To do so, we first establish a formula to express distances between two caterpillar trees in $\rnni$ (\autoref{thm:caterpillar_distance_formula}).
This algorithm can also be used to solve the token swapping problem on lollipop graphs, improving the worst-case running time of the known algorithm \autocite{Kawahara2017-ey}.

\begin{theorem}
	Let $T$ and $R$ be caterpillar trees in $\rnni$ such that \[1 = \rank(a_1)_R = \rank(a_2)_R < \rank(a_3)_R < \ldots < \rank(a_n)_R = n-1\]
	and
	\[P_T := \{(a_i,a_j)\ |\  \rank(a_i)_T < \rank(a_j)_T \text{ and } \rank(a_i)_R > \rank(a_j)_R\},\]
	\begin{align*}
		M_T := &\{a_i\ |\  (\forall l \textnormal{ with } \rank(a_l)_T \leq \rank(a_i)_T: \rank(a_l)_R > \rank(a_i)_R)\} \\
		& \cap \{a_i \ |\  \rank(a_i)_T < \min\{\rank(a_1)_T, \rank(a_2)_T\}\}
	\end{align*}
	Then for ${m_T = |M_T|}$ and ${p_T = |P_T|}$:
	\[d(T,R) = p_T - m_T.\]
	\label{thm:caterpillar_distance_formula}
\end{theorem}
\vspace{-.5cm}

We refer to pairs $(a_i,a_j) \in P_T$, as defined in \autoref{thm:caterpillar_distance_formula}, as transpositions.
The reason for this is that caterpillar trees can be seen as permutations of the set $\{a_1, \ldots, a_n\}$, ordered by the ranks of their parents.
The tree $R$ in the theorem then corresponds to the identity permutation $(a_1, a_2, a_3, \ldots, a_n)$.
Note that there is no one-to-one correspondence between permutations and caterpillar trees.
For example the permutation $(a_2, a_1, a_3, \ldots, a_n)$ corresponds to the tree $R$ as well.
Therefore, the two pairs of leaves sharing their parent in $T$ and $R$, respectively, are not in the set $P_T$.

\begin{proof}
	Let $T$ and $R$ be caterpillar trees as described above and $\widehat d(T,R) := p_T - m_T$.
	For proving $\widehat d(T,R) = d(T,R)$, it is sufficient to show that for all caterpillar trees $T'$ that are neighbour of $T$ it is
	\begin{align}
		\widehat d(T',R) \geq \widehat d(T,R) - 1.
		\label{eq:distance_proof}
	\end{align}
	The fact that inequality (\ref{eq:distance_proof}) implies $\widehat d(T,R) = d(T,R)$ can be established by induction as in \autocite[Theorem 1]{Collienne2020-iu}.

	For proving inequality (\ref{eq:distance_proof}) we first establish $p_{T'} \geq p_T - 1$ and then $m_{T'} \leq m_T + 1$, assuming that $T'$ is a caterpillar tree that is an $\rnni$ neighbour of $T$.
	We then show that $p_{T'} = p_T - 1$ and $m_{T'} = m_T + 1$ cannot both be true simultaneously, which proves inequality (\ref{eq:distance_proof}).

	At most one transposition of $T$ can be resolved in $T'$ because the only move possible between caterpillar trees $T$ and $T'$ is an $\nni$ move exchanging two leaves.
	Hence $p_{T'} \geq p_T - 1$.
	Let $a_k$ and $a_j$ be the leaves that exchange their position between $T$ and $T'$, such that $\rank(a_k)_T < \rank(a_j)_T$.
	Since these are the only leaves that change positions between $T$ and $T'$, they are the only elements that could be in $M_{T'} \setminus M_T$.
	It remains to show $(M_{T'} \setminus M_T) \neq \{a_k, a_j\}$, from which we can conclude that $m_{T'} \leq m_T - 1$.
	We prove this by showing that if $a_k \in (M_{T'} \setminus M_T)$, it follows $a_j \notin M_{T'}$.

	We assume that $a_k \in (M_{T'} \setminus M_T)$, implying $a_k \notin M_T$, so at least one of the following conditions must be violated for $i = k$:
	\setcounter{equation}{0} 
	\renewcommand{\theequation}{C\arabic{equation}}
	\begin{align}
		\forall l \text{ with } \rank(a_l)_T \leq \rank(a_i)_T: \rank(a_l)_R > \rank(a_i)_R \label{condition1}\\
		\rank(a_i)_T < \min\{\rank(a_1)_T, \rank(a_2)_T\}.
		\label{condition2}
	\end{align}
	\setcounter{equation}{1}
	\renewcommand{\theequation}{\arabic{equation}}

	At first we consider the case that (\ref{condition1}) is violated for $a_k$ in $T$.
	Then there is an $l$ such that $\rank(a_l)_T \leq rank(a_k)_T$ and $\rank(a_k)_R > \rank(a_l)_R$.
	It immediately follows that the same condition is violated for $a_k$ in $T'$, because the $\nni$ move exchanging $a_k$ and $a_j$ preserves the relationship of $a_k$ and $a_l$.
	It hence is $a_k \notin M_{T'}$, contradicting our assumption $a_k \in (M_{T'} \setminus M_T)$.

	We can therefore assume that (\ref{condition2}) is violated for $a_k$.
	It follows $\rank(a_k)_T \geq \min\{\rank(a_1)_T, \rank(a_2)_T\}$.
	As only $a_k$ and $a_j$ exchange between $T$ and $T'$ and $a_k \in M_{T'}$, it follows $a_j \in \{a_1, a_2\}$.
	This however results in a violation of (\ref{condition2}) for $a_j$ in $T'$ and hence $a_j \notin M_{T'}$.
	We can conclude $(M_{T'} \setminus M_T) \neq \{a_k, a_j\}$, and hence $m_{T'} \leq m_T + 1$.

	It remains to show that $p_{T'} = p_T - 1$ and $m_{T'} = m_T + 1$ cannot be true at the same time.
	We assume $p_{T'} = p_T - 1$, hence $(a_k,a_j)$ is a transposition in $T$, meaning that $\rank(a_k)_T < \rank(a_j)_T$ and $\rank(a_k)_R > \rank(a_j)_R$.
	As $a_k$ and $a_j$ are the only leaves that could be in $M_{T'} \setminus M_T$, it suffices to show that neither of them actually is in $M_{T'} \setminus M_T$, resulting in $m_{T'} < m_T + 1$.

	That $a_k$ is not in $M_{T'}$ follows from the violation of (\ref{condition1}) by $\rank(a_j)_{T'} < \rank(a_k)_{T'}$ and $\rank(a_j)_R < \rank(a_k)_R$.
	It hence is $a_k \notin M_{T'} \setminus M_T$.
	Moreover, if $a_j \in M_{T'}$, it follows $a_j \in M_T$ as explained in the following.
	If $a_j \in M_{T'}$, both conditions (\ref{condition1}) and (\ref{condition2}) are met by $a_j$ in $T'$.
	With $\rank(a_k)_T < \rank(a_j)_T$ and $\rank(a_k)_R > \rank(a_j)_R$ it immediately follows that these conditions are also met in $T$, and hence $a_j \in M_T$, and therefore $a_j \notin M_{T'} \setminus M_T$.

	Summarising, it is $M_{T'} \setminus M_T = \emptyset$, and we can conclude that if $p_{T'} = p_T - 1$, it is $m_{T'} < m_T + 1$, which concludes this proof.
\end{proof}

\begin{corollary}
	The distance between two caterpillar trees can be computed in $\O(n \sqrt{\log n})$ in $\rnni$.
	\label{cor:caterpillar_distance_rnni_nlogn}
\end{corollary}
\vspace{-0.66cm}

\begin{proof}
	By \autoref{thm:caterpillar_distance_formula} the distance between two caterpillar trees in $\rnni$ is the number of transpositions between two sequences of length $n$ minus $m_T$ as defined in \autoref{thm:caterpillar_distance_formula}.
	The value $m_T$ can be computed in time linear in $n$ for any caterpillar tree $T$ by considering the leaves of the tree ordered according to increasing rank of their parents.
	The number of transpositions of a sequence of length $n$ (Kendall-tau distance) can be computed in time $\O(n \sqrt{\log n})$ \autocite{Chan2010-ls}.
	This number is equal` to $p_T$, as defined in \autoref{thm:caterpillar_distance_formula}, when ignoring transpositions for the pairs of leaves sharing a parent in $T$ and $R$, respectively.
	The worst-case running time for computing the $\rnni$ distance between caterpillar trees is therefore $\O(n \sqrt{\log n})$.
\end{proof}

\subsection{Diameter and Radius}

\label{section:diameter}
\summary{Definition of Diameter.}
In this section we  investigate the \emph{diameter} of $\rnni$ and $\dtt_m$, which is the greatest distance between any pair of trees in each of these graphs, respectively, i.e. $\max\limits_{\text{trees }T,R}d(T,R)$.
We first establish the exact diameter of $\rnni$, improving the upper bound $n^2 - 3n - \frac{5}{8}$ given by \textcite{Gavryushkin2018-ol}.
Afterwards, we generalise this result to $\dtt_m$.

\begin{theorem}
	The diameter of $\rnni$ is $\frac{(n-1)(n-2)}{2}$.
	\label{thm:diameter_rnni}
\end{theorem}

\begin{proof}
	For proving this theorem we use the fact that $\findpath$ computes shortest paths in $\rnni$.
	Each iteration $i$ of $\findpath$, applied to two ranked trees $T$ and $R$, decreases the rank of the most recent common ancestor of a cluster $C$, induced by the node of rank $i$ in $R$, in the currently last tree $T'$ on the already computed path (starting wth $T' = T$).
	The maximum rank of $(C)_{T'}$ at the beginning of iteration $i$ is $n-1$, the rank of the root.
	As every move decreases the rank of $(C)_{T'}$ by one, there are at most $n-1-i$ moves in iteration $i$.
	The maximum length of a shortest path in $\rnni$ is hence $\sum \limits_{i = 1}^{n-1} i = \frac{(n-1)(n-2)}{2}$.
	Note that the caterpillar trees $[\{a_1, a_2\}, \{a_1, a_2, a_3\}, \ldots, \{a_1, \ldots, a_n\}]$ and $[\{a_n, a_{n-1}\}, \{a_n, a_{n-1}, a_{n-2}\}, \ldots, \{a_n, \ldots, a_1\}]$ provide an example of trees that have distance $\frac{(n-1)(n-2)}{2}$, as already pointed out in \autocite[Corollary 1]{Collienne2020-iu}, proving that this upper bound for the length of a shortest path is tight.
\end{proof}

\begin{theorem}
	The diameter of $\dtt_m$ is $\frac{(n-1)(n-2)}{2} + (m-n+1)(n-1)$.
	\label{thm:dtt_diameter}
\end{theorem}

\begin{figure}[ht]
	\includegraphics[width=0.65\textwidth]{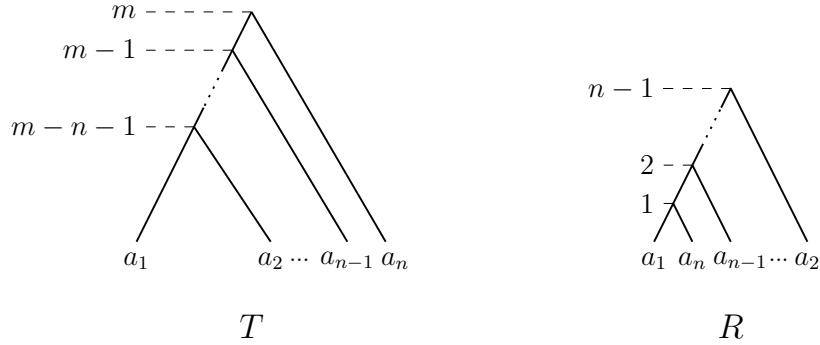}
	\caption{Trees $T$ and $R$ with distance $\frac{(n-1)(n-2)}{2} + (m-n+1)(n-1)$ as described in the proof of \autoref{thm:dtt_diameter}.}
	\label{fig:cat_max_dist_dtt}
\end{figure}

\begin{proof}
	In order to prove the diameter of $\dtt_m$, we consider the longest path that $\findpath$ can compute between any two trees $T$ and $R$.
	That is, we consider the maximum number of moves that $\findpath$ can perform on the extended ranked versions $T_r$ and $R_r$ of any two trees $T$ and $R$.
	Therefore, we distinguish $\rnni$ moves in the subtrees on the leaf set $\{a_1, \ldots, a_n\}$ from the rank moves corresponding to length moves, i.e. rank moves between one node of each of the subtrees on leaf subsets $\{a_1, \ldots, a_n\}$ and $\{a_{n+1}, \ldots a_{m+2}\}$.

	The maximum number of $\rnni$ moves (excluding rank moves corresponding to length moves) on $\fp(T_r,R_r)$ follows from \autoref{thm:diameter_rnni} and is $\frac{(n-1)(n-2)}{2}$.
	The maximum number of rank moves corresponding to length moves on a shortest path between $T_r$ and $R_r$ is reached when every internal node of the subtree $T_r^c$ of $T_r$ swaps rank with every internal node of the subtree $T_r^d$.
	The maximum number of such rank swaps corresponding to length moves is hence $(m-n+1)(n-1)$.

	The sum of the maximum number for $\rnni$ and length moves is therefore $\frac{(n-1)(n-2)}{2} + (m-n+1)(n-1)$.
	To show that this upper bound is actually the diameter of $\dtt_m$ we give an example of trees $T$ and $R$ (\autoref{fig:cat_max_dist_dtt}) for which the path computed by $\findpath$ has length $\frac{(n-1)(n-2)}{2} + (m-n+1)(n-1)$.
	Both $T$ and $R$ are caterpillar trees defined as follows.
	\[m-n-1 = \rank(a_1)_T = \rank(a_2)_T < \rank(a_3)_T < \ldots < \rank(a_n)_T = m\]
	and
	\[1 = \rank(a_1)_R = \rank(a_n)_R < \rank(a_{n-1})_R < \ldots < \rank(a_1)_R = n-1.\]
\end{proof}

Note that the worst-case running time of $\findpath$ is $\O(n^2)$ in $\rnni$ and $\O(nm)$ in $\dtt_m$ and depends on the diameter of the corresponding tree spaces.
For computing a shortest path, there is no algorithm with better worst-case running time than this, as the running time for algorithms computing shortest paths is bounded from below by the diameter of the corresponding space.
There can however be more efficient algorithms for computing distances.
\vspace{12pt}

\summary{Radius of $\rnni$ is equal to its diameter.}
The \emph{radius} of a graph is defined as he minimum distance of any vertex in the graph to the vertex with maximum distance from it, that is, $\min\limits_{T}\max\limits_{R} d(T,R)$, where $d$ is the distance measure in the corresponding graph.
In the following we see that the radius of $\rnni$ equals its diameter, which is not true for $\dtt_m$, as we will see afterwards.

\begin{theorem}
	The radius of $\rnni$ equals its diameter $\frac{(n-1)(n-2)}{2}$.
	\label{thm:radius_rnni}
\end{theorem}

\begin{proof}
	We prove this theorem by showing that every ranked tree $T$ in $\rnni$ has a caterpillar tree $R$ with distance $\frac{(n-1)(n-2)}{2}$ to $T$, using induction on the number of leaves $n$.

	The base case $n=3$ is trivial, as all three trees in this space are caterpillar trees with distance one from each other.
	For the induction step we consider an arbitrary tree $T$ with $n+1$ leaves.
	Let $x$ and $y$ be the leaves of $T$ that share the internal node of rank one as parent in $T$, and let $T_n$ be the tree on $n$ leaves resulting from deleting one of these leaves, say $x$, from $T$, and suppressing the resulting degree-$2$ vertex.
	By the induction hypothesis there is a caterpillar tree $R_n$ with distance $\frac{(n-1)(n-2)}{2}$ to $T_n$.
	Now consider the tree $R$ resulting from adding $x$ at the top of $R_n$ such that the root of $R$ has $x$ and $R_n$ as children.

	We now consider $\fp(R,T)$.
	In the first iteration of $\findpath$, $(\{x,y\})_R$ moves down until it reaches rank one.
	Therefore, first $(x)_R$ moves down by $\nni$ moves until it reaches rank $\rank(y)_R + 1$.
	Then a further $\nni$ move creates an internal node with children $x$ and $y$, before this node is moved down by rank swaps to reach rank one as depicted in Figure~\ref{fig:max_dist_ctree}.
	Altogether, there are $n-1$ $\rnni$ moves needed in the first iteration, as the rank of the parent of $x$ decreases by one within every move, starting at the root with rank $n$ and ending at the internal node of rank one.
	The tree at the end of this first iteration on $\fp(R,T)$ is identical to $R_n$ when removing the leaf $x$ and suppressing its parent (the node of rank one).
	Since the cluster $\{x,y\}$ is not considered again in $\findpath$, the remaining part of $\fp(R,T)$ contains the same moves as $\fp(R_n,T_n)$, and hence $|\fp(R,T)| = |\fp(R_n,R_n)| + n-1$.
	Therefore it is $d(T,R) = \frac{(n-1)(n-2)}{2} + n-1 = \frac{n(n-1)}{2}$, which proves the lemma.
	\begin{figure}[ht]
		\includegraphics[width=0.8\textwidth]{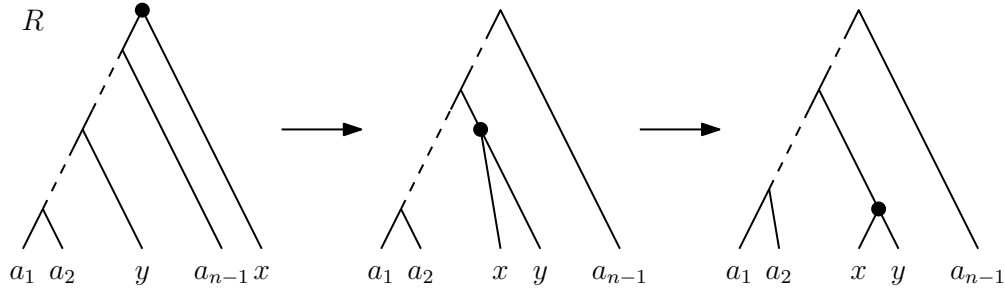}
		\caption{Initial $n - 1$ $\rnni$ moves of $\fp(R,T)$ as described in the proof of \autoref{thm:radius_rnni}.
		Removing the leaf $x$ and suppressing the non-root node of degree two from the tree on the right results in $R_n$ as described in the theorem.}
		\label{fig:max_dist_ctree}
	\end{figure}
\end{proof}

\begin{figure}[ht]
	\includegraphics[width=0.15\textwidth]{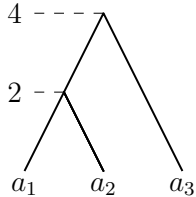}
	\caption{Tree in $\dtt_4$ on three leaves for which there is not tree with distance $5 = \frac{(n-1)(n-2)}{2} + (m-n+1)(n-1)$ (diameter) from it}
	\label{fig:dtt_radius_counterexample}
\end{figure}

Unlike in $\rnni$, the radius of $\dtt_m$ does not equal its diameter.
A counterexample is given by the tree depicted in \autoref{fig:dtt_radius_counterexample} on three leaves in $\dtt_4$.
There is no tree in $\dtt_4$ that has distance $\frac{(n-1)(n-2)}{2} + (m-n+1)(n-1)$ (diameter of $\dtt_m$) from this tree.

\section{Conclusion and future research questions}
\label{section:open_problems}

\summary{Brief summary of results of the paper}
In this paper we introduced and analysed properties of the space of discrete coalescent trees $\dtt_m$.
An important tool for establishing these characteristics of the tree space is the algorithm $\findpath$, which has been introduced by \textcite{Collienne2020-iu} for $\rnni$.
We generalised this algorithm and showed in \autoref{thm:dtt_findpath} that it solves the shortest path problem in $\dtt_m$ as well.
Afterwards, we established properties of $\dtt_m$ and $\rnni$ such as the cluster property (\autoref{section:cluster_property}), the convexity of the set of caterpillar trees (\autoref{section:caterpillar_convex}), diameter, and radius (\autoref{section:diameter}).
With the convexity of the set of caterpillar trees in $\rnni$ we also found a more efficient way of computing distances between such trees, using the correspondence between caterpillar trees and permutations.

\summary{More efficient algorithm for computing distances (not shortest paths)}
The worst-case time complexity of $\findpath$ for computing a shortest path is $\O(mn)$ in $\dtt_m$.
In \autoref{section:diameter} we have seen that there is no algorithm with better worst-case running time for computing shortest paths.
However, it might be possible to compute distances more efficiently.
In fact, we established in \autoref{section:caterpillar_convex} a way for computing distances between caterpillar trees in $\O(n \sqrt{\log n})$.
This raises the question whether there is an algorithm that computes the distance between any two trees in $\dtt_m$ with better running time than $\findpath$.

\summary{$\rnni(rho)$ and parameter $\rho$ for discrete coalescent trees}
Throughout this paper we consider $\dtt_m$ as a generalisation of $\rnni$ by allowing internal nodes to have integer-valued time differences.
We therefore introduced the parameter $m$ to bound the height of a tree in the space of discrete coalescent trees in order to get a finite space.
A different parameter $\rho$ has previously been introduced in \textcite{Collienne2020-iu} for generalising $\rnni$ to a space $\rnni(\rho)$ of ranked trees, where rank and $\nni$ moves have weights $\rho$ and one, respectively.
Combining these two approaches of generalising $\rnni$ gives a space of discrete coalescent trees where different moves have different weights.
This tree space is relevant for practical applications, where for example some knowledge about the tree topology exists, but the uncertainty of the timing of internal nodes remains high.
Investigating such a tree space could therefore be a next step for further studies.

\summary{t-space}
Another tree space, of which $\dtt_m$ is a discretisation, is the $t$-space \autocite{Gavryushkin2016-uu}, where internal nodes are assigned real-valued times.
For this space on time-trees no algorithm for computing shortest paths or distances is known yet.
Our results for $\dtt_m$, however, can be transferred to this space, as discrete coalescent trees can be interpreted as discrete time-trees.
Distances in $\dtt_m$ can therefore be used to approximate those in $t$-space.
For this it is important to notice that the parameter $m$ is not relevant in applications, as distances between two trees in $\dtt_{m'}$ coincide with those in $\dtt_m$ if $m' < m$.
Since choice of $m$, and therefore the choice on how to discretise time-trees, drives the complexity of computing shortest paths (\autoref{section:diameter}), finding a way to discretise time-trees to use our results on $\dtt_m$ can be subject of further research.

\summary{partition lattice -- can be found in supplement}
In \autoref{section:caterpillar_convex} we established a connection between the shortest path problem for caterpillar trees in $\rnni$ and the token swapping problem on lollipop graphs.
We can furthermore provide a connection between the $\rnni$ graph and a well-known algebraic structure, the partition lattice.
We provide a detailed description of this relation in the supplementary material.

\section*{Acknowledgements}
We thank Charles Semple for useful discussions about the Cluster Property, and Mike Steel for his useful comments that improved this paper.

We acknowledge support from the Royal Society Te Ap\=arangi through a Rutherford Discovery Fellowship (RDF-UOO1702).
This work was partially supported by Ministry of Business, Innovation, and Employment of New Zealand through an Endeavour Smart Ideas grant (UOOX1912), a Data Science Programmes grant (UOAX1932).

MF thanks the joint research project \textit{DIG-IT!} supported by the European Social Fund (ESF), reference: ESF/14-BM-A55-0017/19, and the Ministry of Education, Science, and Culture of Mecklenburg-Vorpommern, Germany, for funding parts of this work.

\printbibliography

\section{Supplement}

In the following we discuss a connection of the $\rnni$ graph to a well-known algebraic structure, the partition lattice.
Following that, we discuss further open problems providing new ideas for future research.

\subsection{Partition Lattice}
The connection of $\rnni$ to partition lattices provides a new direction for further research and translates results and open problems from the language of phylogenetics to the language of lattice theory.

\begin{figure}[ht]
\centering
\includegraphics[width=0.9\textwidth]{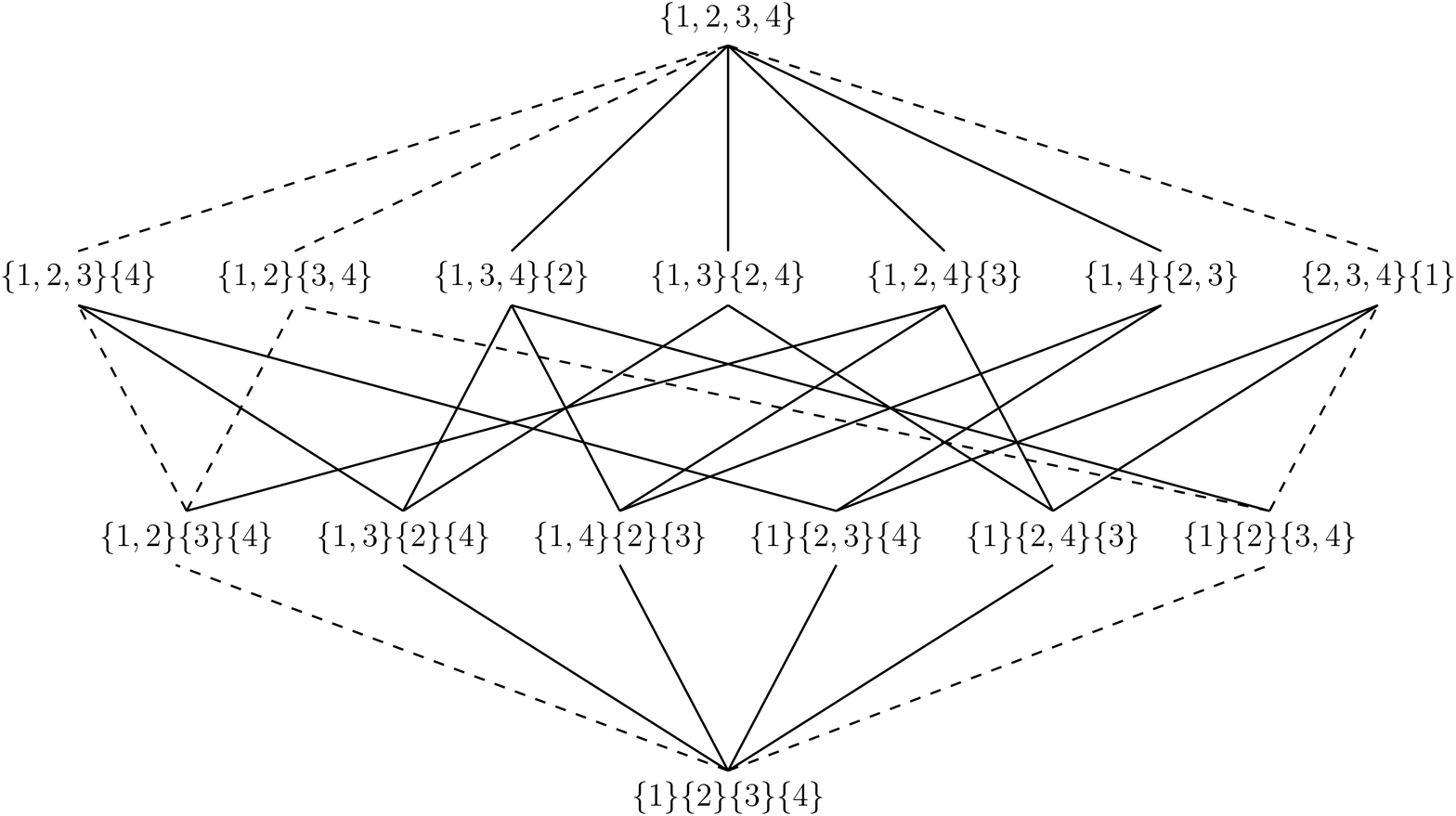}
\vspace{12pt}
\caption{The partition lattice $\Pi_4$ on $\{1,2,3,4\}$.
The dashed edges correspond to an $\rnni$ path from the tree represented by the leftmost chain to the rightmost one.}
\label{fig:partition_lattice4}
\end{figure}

The \emph{partition lattice} on $\{1, \ldots, n\}$ is the lattice given by the partially ordered set $(\Pi_n, \leq)$, where $\Pi_n$ is the power set of $\{1, \ldots, n\}$ and $X \leq Y$ if partition $X$ refines $Y$, that is, $X \leq Y \Leftrightarrow (\forall x \in X)(\exists y \in Y) x \subseteq y$.
A \emph{chain} in a lattice is a set $\{X_0, \ldots, X_k\}$ with $X_0 \leq X_1 \leq \ldots \leq X_k$.
The \emph{length} of the chain $\{X_0, \ldots, X_k\}$ is $k$, the number of its elements minus one, and such a chain is called \emph{maximal chain} of a lattice if there is no chain with length greater than $k$ in the lattice.
For simplification we will denote the partition lattice on $n$ elements by $\Pi_n$.
$\Pi_4$ is illustrated in Figure~\ref{fig:partition_lattice4}.
We assume that a partition $X$ in the partition lattice $\Pi_n$ has rank $k$ if the number of elements in $X$ is $n-k$.
The algebraic structure of $\Pi_n$ is related to the $\rnni$ graph on trees on $n$ leaves in the following way.

\begin{theorem}
The $\rnni$ graph on $n$ leaves is isomorphic to the graph of maximal chains of the partition lattice $\Pi_n$ where two maximal chains are connected by an edge if and only if they differ by exactly one partition.
The corresponding metric spaces are isometric.
\label{thm:partition_lattice}
\end{theorem}

This theorem implies that the algorithms developed for trees, such as $\findpath$, can be used on lattices, and also complexity results from $\rnni$ can be transferred.

\begin{proof}
There is a one-to-one relation between ranked trees and maximum chains in a partition lattice.
We can define a bijective mapping from the set of ranked trees to the set of maximum chains in $\Pi_n$ as follows.
A ranked tree $T$ maps onto a maximum chain $\mathcal{C}_T$ if the set in the partition of rank $i$ in $\mathcal{C}_T$ that is the union of two sets of the partition of rank $i-1$ in $\mathcal{C}_T$ is the cluster induced by the internal node of rank $i$ in $T$.

Note that this bijection is an isomorphism between the $\rnni$ graph and the graph of chains as in the theorem.
Indeed, two chains are different by exactly one partition if and only if the corresponding trees are connected by an $\rnni$ move.
\end{proof}

Figure~\ref{fig:partition_lattice4} is an illustration of the proof of Theorem~\ref{thm:partition_lattice}.
The four chains indicated in bold correspond to the following $\rnni$ path.
The leftmost chain corresponds to the caterpillar tree $[\{1, 2\}, \{1, 2, 3\}, \{1, 2, 3, 4\}]$ (in the cluster representation).
First, the partition $\{1, 2, 3\} \{4\}$ is replaced with $\{1, 2\} \{3, 4\}$ and we get the chain corresponding to the tree $[\{1, 2\}, \{3, 4\}, \{1, 2, 3, 4\}]$, which is one $\rnni$ move away from the caterpillar tree.
Second, the partition $\{1, 2\} \{3\} \{4\}$ is replaced with $\{1\} \{2\} \{3, 4\}$, which corresponds to the rank swap on the previous tree.
Third, the partition $\{1, 2\} \{3, 4\}$ is replaced with $\{1\} \{2, 3, 4\}$ and we reach the caterpillar tree $[\{3,4\}, \{2, 3, 4\}, \{1, 2, 3, 4\}]$.
\end{document}